\documentclass{article}

 \usepackage[preprint]{neurips_2025}

\usepackage[utf8]{inputenc} 
\usepackage[T1]{fontenc}    
\usepackage{hyperref}       
\usepackage{url}            
\usepackage{booktabs}       
\usepackage{amsfonts}       
\usepackage{nicefrac}       
\usepackage{microtype}      
\usepackage{xcolor}         
\usepackage{amsmath} 
\usepackage{arydshln}
\usepackage{graphicx}
\usepackage{algorithmic}
\usepackage{amssymb}
\usepackage{tikz}

\title{FinFlowRL: An Imitation-Reinforcement Learning Framework for Adaptive Stochastic Control in Finance}

\author{%
  Yang Li$^{1}$ \quad
  Zhi Chen$^{1}$ \quad
  Steve Y. Yang$^{1}$ \quad
  Ruixun Zhang$^{2}$ \\
  \\
  $^{1}$School of Business, Stevens Institute of Technology \\
  $^{2}$School of Mathematical Sciences, Peking University \\
  \\
  \texttt{\{yli269, zchen100, syang14\}@stevens.edu} \\
  \texttt{zhangruixun@pku.edu.cn}
}

\begin{document}

\maketitle

\vspace{-5mm}

\begin{abstract}
Traditional stochastic control methods in finance rely on simplifying assumptions that often fail in real-world markets. While these methods work well in specific, well-defined scenarios, they underperform when market conditions change. We introduce \textbf{FinFlowRL}, a novel framework for financial stochastic control that combines imitation learning with reinforcement learning. The framework first pre-trains an adaptive meta-policy by learning from multiple expert strategies, then fine-tunes it through reinforcement learning in the noise space to optimize the generation process. By employing action chunking—generating sequences of actions rather than single decisions—it addresses the non-Markovian nature of financial markets. FinFlowRL consistently outperforms individually optimized experts across diverse market conditions.
\end{abstract}

\vspace{-5mm}
\section{Introduction}




Stochastic control in finance addresses optimal decision-making under uncertainty, fundamental to high-frequency trading, optimal execution \citep{almgren2001optimal}, and portfolio optimization \citep{merton1969lifetime}. Traditional approaches formulate the problem as an objective function governed by a stochastic differential equation (SDE) with constraints. While tractable under stylized assumptions, these models face key limitations in practice. They rely on parameter calibration from historical data, which fails under regime shifts, and often assume stationary dynamics such as geometric Brownian motion \citep{black1973pricing}. Real markets exhibit jumps, stochastic volatility, and non-stationary behavior, making such assumptions unrealistic and solutions suboptimal. Moreover, framing decisions as a Markov Decision Process (MDP) ignores the path-dependent and memory-rich nature of financial markets \citep{gatheral2022volatility}.

We propose FinFlowRL, a two-stage framework for financial stochastic control. First, we pre-train a flow-matching model that learns from multiple expert strategies across diverse market scenarios, integrating their strengths into a unified policy. Second, we fine-tune via reinforcement learning by optimizing the noise generation process rather than actions directly—the pre-trained model remains frozen while we learn to generate better input noise. Both stages employ action chunking \citep{chi2024diffusionpolicy,li2025reinforcement}, generating sequences of decisions over planning horizons rather than single actions, naturally capturing the non-Markovian dynamics of financial markets.

To our knowledge, this is the first framework combining flow matching and RL for financial stochastic control. Applied to high-frequency trading, FinFlowRL consistently outperforms individual expert models and adapts effectively to changing market conditions.

\section{Methodology}

FinFlowRL employs a two-stage approach: (1) pre-training a MeanFlow model based on expert demonstrations, and (2) fine-tuning via reinforcement learning in noise space while keeping the expert frozen. 

\subsection{Stage 1: MeanFlow Pre-training}

\textbf{Expert Demonstration Generation.} We simulate 108 market scenarios varying volatility ($\sigma \in \{0.05, 0.1, 0.3\}$), order arrival rates ($\lambda \in \{10, 20, 40\}$), and jump intensities. For each scenario, we evaluate four experts—Avellaneda-Stoikov (AS) \citep{avellaneda2008high}, GLFT \citep{gueant2013dealing}, modified GLFT with drift \citep{gueant2013dealing}, and PPO \citep{schulman2017proximal}—selecting the best performer's actions as demonstrations, yielding 3.24M state-action pairs.

\textbf{MeanFlow Model.} Unlike standard flow matching that models instantaneous velocity $v(z_t, t)$, MeanFlow \citep{geng2025mean} models average velocity between time steps:
$$u(z_t, r, t) = \frac{1}{t-r} \int_{r}^{t}v(z_\tau, \tau)d\tau$$

The key insight is the MeanFlow Identity, which connects average and instantaneous velocities:
$$u(z_t, r, t) = v(z_t, t) - (t - r)\frac{d}{dt}u(z_t, r, t)$$

This relationship enables training without access to the true instantaneous velocity. During training, we construct interpolants between noise $z_0 \sim \mathcal{N}(0, I)$ and expert actions $a_{expert}$:
$$z_t = (1-t)z_0 + t \cdot a_{expert}$$

The training objective minimizes:
$$\mathcal{L}(\theta) = \mathbb{E}_{t,r,z_t,s} \left[ \| u_\theta(z_t, r, t, s) - \text{sg}(u_{target}) \|^2 \right]$$
where $u_{target} = v_t - (t-r)(v_t\partial_z u_\theta + \partial_t u_\theta)$ and $v_t = a_{expert} - z_0$ is the straight-line velocity.

\textbf{Conditioning and Generation.} We condition on market state $s$ using FiLM \citep{perez2018film}, which modulates network features via $\mathbf{h}' = \boldsymbol{\gamma}(s) \odot \mathbf{h} + \boldsymbol{\beta}(s)$. This enables state-dependent action generation through one-step inference:
$$a = z_1 - u_\theta(z_1, 0, 1, s)$$
where we set $r=0, t=1$ for generation. This one-step formulation is crucial for meeting HFT's microsecond latency requirements.

\subsection{Stage 2: FlowRL Fine-tuning}

Instead of retraining the entire model, we freeze the pre-trained MeanFlow expert $g_\theta$ and learn only a noise policy $\pi^W_\phi$ that generates optimal input noise. Following \citep{lv2025flow, wagenmaker2025steering}, we transform the MDP from action space to noise space:

$$w \sim \pi^W_\phi(\cdot|s), \quad a = g_\theta(s, w)$$

The noise policy is a Gaussian $\pi^W_\phi(w|s) = \mathcal{N}(\mu_\phi(s), \Sigma_\phi)$ optimized via PPO with clipped objective:
$$L^{PPO} = \mathbb{E}_t[\min(r_t(\theta)A_t, \text{clip}(r_t(\theta), 1-\epsilon, 1+\epsilon)A_t)]$$

This approach reduces trainable parameters by 84\% while leveraging pre-trained knowledge.

\subsection{Action Chunking}
Both stages generate action sequences rather than single decisions. Given observation window $T_{obs}=2$, we predict $T_{pred}=8$ future actions but execute only $T_{exec}=4$. This hierarchical planning captures temporal dependencies and market memory, directly addressing non-Markovian dynamics inherent in financial markets.

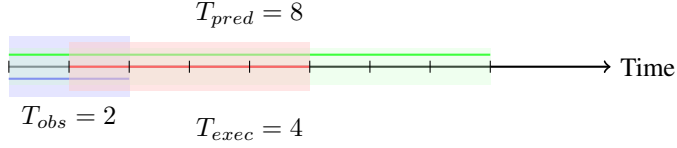
\begin{figure}[h]
\centering
\begin{tikzpicture}[scale=0.8]
    \draw[thick, ->] (0,0) -- (10,0) node[right] {Time};
    
    \draw[thick, blue] (0,-0.2) -- (2,-0.2);
    \fill[blue!20, opacity=0.5] (0,-0.5) rectangle (2,0.5);
    \node[below] at (1,-0.5) {$T_{obs}=2$};
    
    \draw[thick, green] (0,0.2) -- (8,0.2);
    \fill[green!20, opacity=0.3] (0,-0.3) rectangle (8,0.3);
    \node[above] at (4,0.5) {$T_{pred}=8$};
    
    \draw[thick, red] (1,0) -- (5,0);
    \fill[red!20, opacity=0.5] (1,-0.4) rectangle (5,0.4);
    \node[below] at (4,-0.7) {$T_{exec}=4$};
    
    \foreach \i in {0,...,8} {
        \draw (\i,0.1) -- (\i,-0.1);
    }
\end{tikzpicture}
\caption{Hierarchical temporal structure in MeanFlow-PPO. The model observes $T_{obs}$ past states, generates actions for $T_{pred}$ future steps, but only executes the first $T_{exec}$ actions before replanning.}
\label{fig:temporal_structure}
\end{figure}

\section{Application in High-Frequency Trading}

\subsection{Problem Formulation}
We formulate high-frequency trading market-making as a stochastic control problem over discrete time steps $t \in \{0, \dots, T\}$. At each step, an agent observes state $O_t$ (including market data and private information like inventory) and takes action $A_t$ (typically setting bid and ask spreads $(\delta^b_t, \delta^a_t)$).

The goal is to learn an optimal policy $\pi(O_t) = A_t$ that maximizes expected terminal wealth minus inventory risk:
$$\max_{\pi} \mathbb{E}^{\pi} \left[ W_T - \phi(I_T) \mid O_0 \right]$$
where $W_T$ is terminal wealth and $\phi(I_T)$ penalizes unsold inventory.

\subsection{Generating Market Observation-Action Pairs}
We model mid-price $S_t$ as a jump-diffusion process \citep{merton1976option}:
$$dS_t = S_{t^-} \left( \mu dt + \sigma dB_H(t) \right) + S_{t^-} (e^{J} - 1) dN_t$$
where $\mu$ is drift, $\sigma$ is volatility, $dB_H(t)$ is fractional Brownian motion \citep{mandelbrot1968fractional}, $J \sim N(\mu_J, \sigma_J^2)$ represents jump size, and $dN_t$ is a Poisson process with intensity $\lambda_J$.

Order arrivals follow mutually exciting Hawkes processes \citep{bacry2015hawkes,hawkes1971point}, capturing self-exciting (previous buy/sell orders increase subsequent same-type arrivals) and cross-exciting effects (buy orders influence sell arrivals and vice versa).

Buy and sell order intensities are:
\begin{align}
\lambda_a(t) &= \mu_a + \sum_{t_i \in \mathcal{N}_a} \alpha_{aa} e^{-\beta (t - t_i)} + \sum_{t_j \in \mathcal{N}_b} \alpha_{ab} e^{-\beta (t - t_j)}\\
\lambda_b(t) &= \mu_b + \sum_{t_i \in \mathcal{N}_b} \alpha_{bb} e^{-\beta (t - t_i)} + \sum_{t_j \in \mathcal{N}_a} \alpha_{ba} e^{-\beta (t - t_j)}
\end{align}

We create market scenarios with varying liquidity levels (high, medium, low) and stress conditions featuring sudden changes and increased volatility.

Our expert candidates include: (1) Avellaneda-Stoikov (AS) model, (2) Guéant-Lehalle-Fernandez-Tapia (GLFT) model, (3) modified GLFT with price drift, and (4) PPO-trained RL agent \citep{schulman2017proximal}. We generate actions from each agent for each market scenario, collecting 3.24 million state-action pairs.

\subsection{Results}
Our investigation is designed to critically evaluate FinFlowRL's advantages over existing approaches and its capabilities to earn profit. 
Specifically, we address the following research questions:
\textbf{RQ1:} Can FinFlowRL effectively generalize strategies learned from expert demonstrations to new, unseen market conditions? \textbf{RQ2:} Does the fine-tuning mechanism improve the performance of actions initially proposed by the pre-trained model? \textbf{RQ3:} Can the FinFlowRL framework achieve greater profitability than traditional strategies?

To evaluate FinFlowRL's performance across a spectrum of out-of-sample market environments, we systematically configured distinct test conditions by setting key market parameters to differ from those used during training. We differentiated the market microstructure by creating four specific situations based on combinations of market volatility (Vol) and overall order arrival rate (AR): (1) High Vol/High AR (HH), representing active, potentially news-driven volatile markets; (2) High Vol/Low AR (HL), mimicking risky market; (3) Low Vol/High AR (LH), reflecting stable, liquid markets; and (4) Low Vol/Low AR (LL), simulating quiet market periods. For each of these market situations, we perform traditional strategies as well as FinFlowRL to compare their performance.

We take the following evaluation metrics.
(1) Profit and Loss (PnL). Measure the total change in the value over a specific period, reflecting the aggregate percentage gain or loss.
(2) Sharpe Ratio (SR). It helps investors understand how much excess return an investment generated for each unit of risk it undertook. A higher Sharpe Ratio generally indicates a better risk-adjusted performance. 
(3) Maximum Drawdown (MDD) Quantify the largest percentage decline in the value of a strategy from a previous peak to a subsequent trough.

\begin{table}[!htbp]
\centering
\caption{Performance comparison across market conditions. Parameters: Hurst Exponent $H=0.5$, Drift Rate $\mu=0$, volatility ($\sigma$: \{0.02, 0.25\}) and arrival rate ($\lambda$: \{25, 50\}). Each method evaluated over 1 million trials. PnL: Profit and Loss, SR: Sharpe Ratio, MDD: Maximum Drawdown (\%).}
\label{tab:performance}
\renewcommand{\arraystretch}{1.2}
\resizebox{\textwidth}{!}{%
\begin{tabular}{lrrr|rrr|rrr|rrr}
\toprule
& \multicolumn{3}{c|}{High Volatility \& High Demand} & \multicolumn{3}{c|}{High Volatility \& Low Demand} & \multicolumn{3}{c|}{Low Volatility \& High Demand} & \multicolumn{3}{c}{Low Volatility \& Low Demand} \\
& PnL $\uparrow$ & SR $\uparrow$ & MDD $\downarrow$ & PnL $\uparrow$ & SR $\uparrow$ & MDD $\downarrow$ & PnL $\uparrow$ & SR $\uparrow$ & MDD $\downarrow$ & PnL $\uparrow$ & SR $\uparrow$ & MDD $\downarrow$ \\
\midrule
Random Action & 1.99& 0.06& 28.49& 0.99& 0.04& 19.24& 2.10 & 0.31 & 2.71 & 1.08 & 0.22 & 1.87 \\
AS & 24.22& 0.09& 241.65& 13.54& 0.09& 125.78& 25.20 & 1.05 & 7.66 & 13.67 & 0.72 & 6.61 \\
GLFT & 25.10& 0.37& 60.57& 13.56& 0.24& 52.55& 25.87 & 1.17 & 6.95 & 13.91 & 0.78 & 6.14 \\
GLFT-drift & 25.10& 0.37& 60.57& 13.56& 0.24& 52.55& 25.87 & 1.17 & 6.95 & 13.91 & 0.78 & 6.14 \\
Vanilla PPO & 14.76& 0.10& 133.61& 9.29& 0.08& 103.85& 26.74 & 0.81 & 10.13 & 19.80 & 0.46 & 14.56 \\
Pretrained MeanFlow  & 23.91& 0.37& 43.4& 12.97& 0.22& 45.47& 23.82& 1.83& 2.18& 12.93& 1.07& 2.69\\
\textbf{FinFlowRL} & 26.33& 0.50& 45.47& 14.32& 0.28& 45.35& 26.27& 2.34& 2.68& 14.29& 1.36& 3.08\\
\bottomrule
\end{tabular}%
}
\end{table}

Table \ref{tab:performance} presents comprehensive results based on 1 million evaluation trials per method to ensure statistical significance. 
Across all tested market conditions, from high to low volatility, \textsc{FinFlowRL} consistently demonstrates superior performance. It achieves the highest risk-adjusted returns (Sharpe Ratios) and maintains lower maximum drawdowns compared to both traditional models and standard reinforcement learning (PPO) approaches.
For \textbf{RQ1}, the initial pre-trained model successfully learns from the collection of experts, achieving performance comparable to its best instructor (the GLFT model). This proves the imitation learning stage effectively internalizes expert strategies. For \textbf{RQ2}, The fine-tuned model significantly outperforms all baseline methods, including the experts it learned from and the initial pre-trained model. This demonstrates that the reinforcement learning stage successfully improves upon the expert knowledge to discover superior, adaptive strategies tailored to specific market conditions. For \textbf{RQ3}, our model consistently achieves the highest profit. \textsc{FinFlowRL} shows remarkable stability, especially in volatile markets with sudden price jumps. This is attributed to its use of "action chunking"—generating sequences of actions—which provides a longer-term planning perspective and mitigates the risk of compounding errors.

\section{Conclusions}

We introduced FinFlowRL, a novel framework for financial stochastic control that combines imitation learning with reinforcement learning. The framework first learns from multiple expert strategies through a MeanFlow policy, then fine-tunes by optimizing noise generation rather than actions directly.

Our main contributions include: (1) first application of flow matching to financial stochastic control, achieving  higher Sharpe ratios and  lower maximum drawdowns than traditional methods; (2) efficient architecture using frozen experts with learnable noise policies, reducing trainable parameters by 84\%; (3) action chunking mechanism that addresses non-Markovian market dynamics

The simulation results demonstrate FinFlowRL's superiority across diverse market conditions. The pre-trained model successfully learns expert strategies, while fine-tuning discovers policies that outperform individual experts. Action chunking proves particularly effective during market jumps, mitigating error compounding inherent in single-step decisions.

FinFlowRL demonstrates that combining expert knowledge with adaptive learning overcomes limitations of purely model-based or data-driven approaches, offering a practical solution for complex financial stochastic control problems.

\bibliographystyle{apalike}
\bibliography{references}










\end{document}